\documentclass{ws-ijmpe}
\usepackage[super,compress]{cite}

\def\bi{\bibitem}

\def\la{\langle}\def\ra{\rangle}
\def\be{\begin{eqnarray}}\def\ee{\end{eqnarray}}
\def\lsim{\mathrel{\rlap{\lower3pt\hbox{\hskip1pt$\sim$}}
		\raise1pt\hbox{$<$}}} 
\def\gsim{\mathrel{\rlap{\lower3pt\hbox{\hskip1pt$\sim$}}
		\raise1pt\hbox{$>$}}} 

\begin{document}

\markboth{Mannque Rho and Long-Qi Shao}{$V_{lowk}$ RG, VM and Sound Velocity in Massive Compact Stars}

\catchline{}{}{}{}{}

\title{$\mathbf{V_{lowk}}$ Renormalization Group Flow, Vector Manifestation\\ and Sound Velocity in Massive Compact Stars}    

\author{Mannque Rho}
\address{\it Universit\'e Paris-Saclay, CNRS, CEA, Institut de Physique Th\'eorique, \\ 91191,   Gif-sur-Yvette, France 
\\ mannque.rho@ipht.fr}

\author{Long-Qi Shao}
\address{School of Fundamental Physics and Mathematical Sciences, \\ Hangzhou Institute for Advanced Study, UCAS, Hangzhou, 310024, China
 \\ shaolongqi22@mails.ucas.ac.cn}

\maketitle

\begin{history}
\received{Day Month Year}
\revised{Day Month Year}
\end{history}

\begin{abstract}
 The $V_{lowk}$-renormalization group approach on the surface of  Fermi liquid for nuclear matter to which Tom Kuo made a pioneering contribution at Stony Brook  is found to inject  the pivotal input  in the formulation of the  generalized nuclear effective field theory with acronym  ``G$n$EFT" applicable to superdense compact-star physics.   A topology change in terms of skyrmions and half-skyrmions is shown to play the role of  the ``putative" hadron-quark continuity (HQC)" conjectured in QCD.  Crucially involved are hidden local symmetry (``HLS") and hidden scale symmetry (``HSS") with the vacuum sliding with density in nuclear medium, with the nuclear tensor force emerging as a Landau Fermi-liquid fixed-point quantity. A possibly novel paradigm, a ``Cheshire Catism,"  in nuclear correlations is suggested.
\end{abstract}


\ccode{PACS numbers:}


\section{Introduction}
Highly dense nuclear matter most likely present in the interior of compact stars of mass $M\gsim 2 M_\odot$ is presently unaccessible directly by QCD, the nonabelian gauge theory of the strong interactions in the Standard Model. Therefore it is currently not feasible to describe in a well-controlled and systematic theoretical framework the density regime relevant to compact stars ranging from the normal nuclear matter density $n_0\sim 0.16$ fm$^{-3}$ to the possible central density of massive stars $n_{cs}\sim (4-7) n_0$ and confront the observations from the past and on-going gravity-wave measurements. This conundrum has generated a huge activity in astrophysical, nuclear and particle physics research (for a recent review focused on the nuclear EoS, among many in the literature,  see  \cite{lattimer}). The relevant high density regime is currently  inaccessible by QCD on lattice (in contrast to high temperature), the effective field theory (EFT) approach based on chiral symmetry ($\chi$EFT for short) breaks down at a density 2 or 3 times $n_0$ and the perturbative QCD anchored on asymptotic freedom (pQCD),  applicable at some high density,  cannot be reliably brought down to the density where the $\chi$EFT breaks down. This essentially makes the compact-star density regime a  theoretical wilderness, a ``jungle."
\vskip 0.2cm
$\bullet$ {\bf EoS ``architecture"}:
Now the state of the matter inside a massive compact star, stable but on the verge of collapse into a black hole,  is a basic physics issue,  a challenge not just to nuclear physics but more fundamentally to QCD per se even if the theory of gravitation is assumed to be well under control. As befits a ``jungle," there are a ``humongous" number of models built in nuclear many-body approaches to cover the full density range  from very low density to very high density relevant to the structure of massive compact stars. Depending on how many parameters are invoked, one may be able to arrive more or less adequately at fitting the available astrophysical and terrestrial laboratory observations by combining the chiral-symmetric nuclear effective field theory \` a la Weinberg ($\chi$EFT$_\pi$ for short) resorting to the pion field only and the density functional approach \`a  la Hohenberg-Kohn theorem (DFA for short) with (phenomenological) multi-scalar meson fields ($s$) and multi-vector meson fields (${\cal V}_\mu$) to account for  from low to higher density regime relevant to the structure of massive compact stars. The $\chi$EFT$_\pi$ is found to work fairly well  up to $n\lsim 2 n_0$ in the equation of state (EoS) when the power expansion is done to order N$^\kappa$ to the leading chiral order ( N$^\kappa$LO) for $\kappa\lsim 4$ but it -- by premise of the EFT -- is to break down at higher densities. The DFA in the form of relativistic mean-field (RMF) approach matching by fiat,  in accuracy,  $\chi$EFT$_\pi$ up to its breakdown density $n_{\chi BD}$ can be tweaked to higher densities by suitable readjustments of the parameters involved, the number of such parameters inevitably increasing as the density is increased.  For convenience, let us call the resulting theory, comprised of both the $\chi$EFT$_\pi$ and DFA,  ``standard nuclear EFT (S$n$EFT)."  It may not be impossible to achieve a reasonable agreement with whatever data available by adjusting arbitrarily large number of free parameters in the S$n$EFT. It makes sense then to ask what the physics is captured by the parameter changes required for the fit.  For this to be feasible, it would be indispensable to reconstruct from potentially accurate neutron-star observational data as well as terrestrial laboratory measurements the ``architecture" of the equation of state (EoS) of baryonic matter without relying on uncertain/unreliable  theoretical inputs or prejudices. There are currently numerous ongoing efforts to set up strategies to map the architecture that will be built from the next generation of astrophysical  observations to realistic EoSs. To give an instance among many,  the potential of such an artificial neural network -- a ``deep learning" approach -- has been discussed~\cite{krastev-MDPI}.
\vskip 0.2cm
$\bullet$ {\bf Density ``ladder"}:
%
Suppose that such an accurate ``experimental" EoS is available. Then starting from the low density regime where a reliable S$n$EFT is available, one may go up in density by ``building an EoS density ladder"~\cite{SP-MDPI}. To start with, nonlinear meson fields with suitable quantum numbers incorporated, one calibrates the parameters to the ground states of finite nuclei and infinite matter supported by $\chi$EFT$_\pi$ near the density regime $\sim n_0$ and then go up the ladders in relativistic (or nonrelativistic) mean-field approaches by recalibrating the parameters  of the density functional in consistency with observational constraints given by the ``architecture" constructed with ``accurate" terrestrial observations -- such as EM probes and low-T heavy ions -- and astrophysical ones. In doing this,  a Bayesian framework is, typically, adopted~\cite{SP-MDPI}. 

Now the question is what is the structure of the so-deduced covariant density functional (CDF) that is consistent with the constraints? 
\vskip 0.2cm

$\bullet$ {$\mathbf{V_{lowk}}$ {\bf as CDF}}:
 At the simplest level of CDF, Walecka's linear relativistic mean-field model with the nucleon, scalar $s$, and iso-scalar vector $\omega$  fields put in as phenomenological fields gave too big a compression modulus $K$. This problem was remedied by introducing multi-field operators with suitable parameters in consistency with known global symmetries of nuclear dynamics. With a limited number of c-number parameters of the phenomenological Lagrangian adjusted so as to fit the ladder of nuclear properties, the properties of neutron stars could be addressed up to certain densities.  That the Walecka linear relativistic field scheme gave rise to Landau Fermi-liquid theory in nuclear matter was already noticed a long time ago by Matsui~\cite{matsui}.  As will be elaborated below,  what is captured in the mean-field treatment of generic CDF is roughly equivalent to a renormalization-group (RG) approach to interacting fermions on the Fermi surface~\cite{shankar,polchinski}. It has been extended to the $V_{lowk}$-RG approach in nuclear physics~{\cite{bogner}.\footnote{A brief history of how the Shankar-Polchinski-RG  (SPRG) approach  entered in nuclear physics in connection with what's treated in this article might be appropriate. Soon after the appearance of the SPRG in journals in $\sim$ 1992-1994, a seminar on this approach in nuclear physics was presented by MR at Stony Brook at which both Gerry Brown and Tom Kuo, together with their PhD students, were present. Combining this development with the BR-scaled hidden local symmetric effective Lagrangian, the Fermi-liquid fixed point  approximation -- that turns out to play the key role in what developed subsequently -- was then formulated in $\sim 1996$  by Bengt Friman and MR addressing nuclear EW processes. Putting the same strategy to finite nuclear systems culminated in the $V_{lowk}$ formalism by Brown, Kuo and their students in $\sim 2001$, which is the tool exploited in what follows.}  This will be developed in nuclear medium with the incorporation of the symmetries, both local and scale, hidden in QCD in place of the phenomenological meson fields figuring in covariant density functional to confront dense baryonic matter currently probed by gravity-wave signals. 
 
 It is here that Tom Kuo brought the $V_{lowk}$-RG approach into the physics of dense baryonic matter dubbed as  ``pseudo-conformal model (PCM)."  
 
 As described below   it will involve a ``double-decimation" RG on the Fermi surface with the first decimation leading to the Fermi-liquid fixed points for the Landau parameters $F$ with fixed  $k_F$ from an effective field theory of QCD and the fluctuations emerging as $1/\bar{N}$ corrections (where $\bar{N}=k_f/[\Lambda_{fs}-k_F])$  in going up the density ladder. At the density regime at which low-energy S$n$EFT breaks down,  there will be a putative ``hadron-quark continuity (HQC)" expected with the onset of what corresponds to the QCD (quark-gluon) degrees of freedom. In this connection, it will be found that a highly intricate role of Brown-Rho (BR) scaling is found to assure the thermodynamic consistency in the CDF-type equations. In this paper an extremely simplified but coherent unique field theory model to which Tom Kuo made a seminal contribution in the $V_{lowk}$ approach he developed at Stony Brook is presented. It has the advantage over other approaches in the literature in that its extremely simple prediction could be unambiguously tested -- either right or wrong -- by future (hopefully) precision astrophysical as well as terrestrial laboratory observables.
 \section{G$\mathbf{n}$EFT}
 The effective theory that enables one to go beyond the S$n$EFT with $\chi$EFT$_\pi$ Lagrangian (and the CDF) in stepping up the density ladder, was dubbed ``generalized nuclear effective field theory (G$n$EFT)." 
 The underlying point of view was that the process going from the nuclear matter density to the massive compact-star densities on the verge of collapse into black holes may not necessitate bona-fide phase transitions in the sense of the Landau-Ginzburg-Wilson paradigm. A plausible scenario that was -- and will be adopted here -- is extensively reviewed in the literature, so we will be as concise as possible. We will mainly define the terminologies used to develop the main ideas. (For appropriate reviews with ample references with the focus on compact-star physics, see  \cite{PCM}.)
 
 There are two key ingredients involved for the development.  The first is the symmetries, hidden in QCD, that are incorporated in the most economical form in the EFT Lagrangian. The  hidden symmetry fields absent in S$n$EFT are endowed with possibly -- non-supersymmetric --``Seiberg-dual"  symmetry structure to QCD, distinguished from phenomenological fields figuring  in the CDF approach. Second  the putative hadron-quark continuity is to be implemented with topology change without involving intrinsic QCD variables. 
 
The first,  hidden symmetries,   is to bring the scale of the degrees of freedom involved from the cutoff $\Lambda_{eft} < m_V$ applicable for $\chi$EFT$_\pi$ (for S$n$EFT) to $\gsim m_V$ where $m_V$ is the vector meson mass $\gsim 3 m_\pi$. The increase of the cutoff is to account for the putative HQC at a density $2 \lsim n/n_0 \lsim 4$ at which $\chi$EFT is expected to break down.   We take into account two symmetries, one local and the other scale/conformal\footnote{We will not distinguish the difference between them -- if any -- in what we are concerned with.}~\cite{HRW}.   The former is the hidden local symmetry (HLS) with the $U(2)=SU(2)\times U(1)$ vector fields ${\cal V}_\mu =\rho_\mu, \omega_\mu$~\cite{HLS,HY-HLS,HLS-grassmannian}.\footnote{Unless otherwise mentioned, we will be limited to two flavors, $N_f=2$.}  Now the HLS is gauge-equivalent to the nonlinear sigma model, so chiral symmetry is encoded therein.\footnote{For future development, an infinite tower of vector fields could be considered in the context of holographic dual QCD which seems to figure in the problem. This is an open problem for the future.}. 

The latter is hidden scale symmetry (HSS) involving a (scalar) dilaton that we denote as $\sigma_d$, not to be confused with the scalar $\sigma$ in the linear sigma model,  that could be associated with $f_0(500)$ listed in the particle data booklet. There has been a long controversy on this since 1970s, more recently in connection with dilatonic Higgs boson in going BSM, for and against an IR fixed point associated with $\sigma_d$ in QCD for $N_f \lsim 3$ (relevant to nuclear physics). We adopt the recent proposal that there can be present an IR fixed point in the confinement regime of QCD, with chiral and scale symmetries spontaneously broken with Nambu-Goldstone bosons, $\pi$ and  dilaton $\sigma_d$, co-existing with massive hadrons, nucleons $\psi$, vector mesons ${\cal V}_\mu$ etc.\footnote{The existence of an IR fixed point in QCD with small $N_f$, say,  $ \sim  3$ is strongly disputed in \cite{yamawaki}. We will suggest that this point is most likely not relevant in our case where the scale symmetry involved is assumed -- and -- expected to be emergent driven by  nuclear correlations.} There are two versions with considerable overlaps, one what's called ``Genuine Dilaton (GD)"~\cite{GD} and the other called ``Conformal Dilaton (CD)-QCD"~\cite{zwicky,beta*,zwicky-pions}.  
 
The effective Lagrangian from which we construct the G$n$EFT  is denoted as $\cal{L}_{\psi {\cal V} \chi}$ and will be  broadly referred to as scale-chiral Lagrangian. In this Lagrangian,   $\psi$ represents the baryon fields, ${\cal V}$ the vector fields and $\chi=f_\chi e^{\sigma_d/f_\chi}$ for the ``conformal compensator" field.   It should be stressed that the Lagrangian $\cal{L}_{\psi {\cal V} \chi}$ is constructed~\cite{PCM} in a way basically different from S$n$EFT.  Embedded in nuclear medium, the parameters of the Lagrangian are endowed with the BR-scaling~\cite{BR91} due to the dilaton condensate $\la\chi\ra^\ast$ sliding with density with the $\ast$ representing the density dependence. Now the density dependence has to be made thermodynamically consistent~\cite{thermodynamic,Song}.\footnote{The naive density dependence  with the c-number $n$ in the Lagrangian parameters violates thermodynamic consistency.  The strategy applied in \cite{Song} was to replace the density $n$ by $\hat{n}$ where $\hat{n}u^\mu\equiv\bar{\psi}\gamma^\mu\psi$ with $u^\mu$ the unit four velocity. This effectively takes into account of the  ``rearrangement terms" simulating multi-field effects in the covariant density formalism. It may be feasible to make the strategy more powerful.} This consistency renders the mean-field treatment of $\cal{L}_{\psi {\cal V} \chi}$ at the leading scale-chiral order equivalent  to the Fermi-liquid fixed-point approximation~\cite{shankar,polchinski}  formulated from a nuclear chiral EFT~\cite{FR,Song}. It can be interpreted as an improvement of a much elaborated covariant energy density functional with however a lot fewer adjustable  parameters than what enters in the  CDF with multi phenomenological meson fields. Up to nuclear matter density $n_0$, its prediction~\cite{PCM} is comparable in precision to  the ground-state properties given in S$n$EFT including the compression modulus, the symmetry energy etc.
\section{Topology for hadron-quark continuity}
What makes the G$n$EFT access higher densities above $\gsim 2 n_0$ is a topological ``twist" to the BR scaling~\cite{topological-twist}. We proposed a topology change  to capture the  putative hadron-quark continuity (HQC) at  density $n_{hqc} \gsim 2 n_0$. This changeover is not visible in terms of baryon fields but it can be in terms of skyrmions. This is because the skyrmion description of baryonic matter at high density becomes more reliable for  large $N_c$. Now  how can this phenomenon be implemented in G$n$EFT? The answer to this question lies in how the hidden symmetries manifest before and after the crossover density $n_{hqc}$.

Putting skyrmion matter on crystal lattice considered to describe QCD more reliably at high density in the large $N_c$ limit,  the high density phase was found to support skyrmion-half-skyrmion transition~\cite{dongetal,PKLR}\footnote{It is with these papers that Tom Kuo entered in the collaboration where his work on $V_{lowk}$-RG makes the pivotal contribution to the structure of G$n$EFT in compact stars.}.  The density at which this takes place is denoted in the literature as $n_{1/2}$. It involves the quark condensate $\la\bar{q}q\ra$ going to zero when averaged on the lattice sites while the pion decay constant $f_\pi$ remaining non-vanishing.  Therefore while the transition  involves topology change, there is no change in chiral symmetry, the order parameter being the quark condensate $\la\bar{q}q\ra$, hence  it is not a ``phase transition" in the sense of the GLW paradigm. What takes place is akin to the pseudo-gap phenomenon in hight-T superconductivity. On the skyrmion lattice in the absence of the hidden symmetries, this changeover appears as a cusp in the symmetry energy  $E_{sym}$ as an $O(1/N_c)$ term in the single-nucleon energy proportional to $\alpha=(N-P)/(N+P)$ where $N(P)$ is the number of neutrons (protons). The cusp is however smoothed into an inflection when the hidden symmetry degrees of freedom are injected.

Now this changeover of the state of the matter with no phase transitions, translated to the behavior of the parameters of the Lagrangian $\cal{L}_{\psi {\cal V} \chi}$ which is treated at the mean-field level (that is in the Landau Fermi-liquid fixed point approximation~\cite{FR}), gives rise to the change from soft to hard equation of state encoded in the inflection, e.g., through the symmetry energy $E_{sym}$ at $n_{1/2}$. It is found compatible, so far with no tension with nature, with what accounts for HQC at $n_{hqc}$ in massive compact stars~\cite{WCU}. This suggests identifying $n_{hqc}$ with  $n_{1/2}$. We adopted this identification. This leads to the first important observation.
\vskip 0.3cm
$\bullet$ {\bf  Skyrmion-to-half-skyrmion transition plays the role of hadron-quark crossover.}  The density range  for the transition is found to be $n_{1/2}\approx n_{hqc} \sim (2-4)n_0$.

What is  crucial in what takes place at $n_{1/2}$ is the structure of the nuclear tensor force given by the exchange of one pion and one $\rho$ meson given by the Lagrangian $\cal{L}_{\psi {\cal V} \chi}$ with the masses and coupling constants BR-scaling with density. With increasing density, the two components of the tensor force tend to cancel. This property of the net tensor force was most remarkably confirmed in the long life-time of $^{14}$C beta decay~\cite{C14}\footnote{Resorting to chiral three-body potential leads to the same result. The physics is equivalent in the sense of G$n$EFT.}.  It is not straightforward to see in the intricate interplay between the spectrum and transition matrix elements how the tensor force affects the $^{14}$C decay but it was pointed out to the author in private communication by Tom Kuo~\cite{pristine} that  when formulated on the Fermi surface, the  $V^{\rm tensor}_{lowk}$-RG~\cite{VlowkFS}  corresponds to the Landau Fermi-liquid fixed point quantity one gets in G$n$EFT for a given $k_F$ as shown in \cite{FR}.  This is convincingly verified in nuclei in the pf-shell region~\cite{otsuka} as shown in Fig.~\ref{pfshell}.
\begin{figure}[htb]
\begin{center}
  \includegraphics[width=10cm,clip]{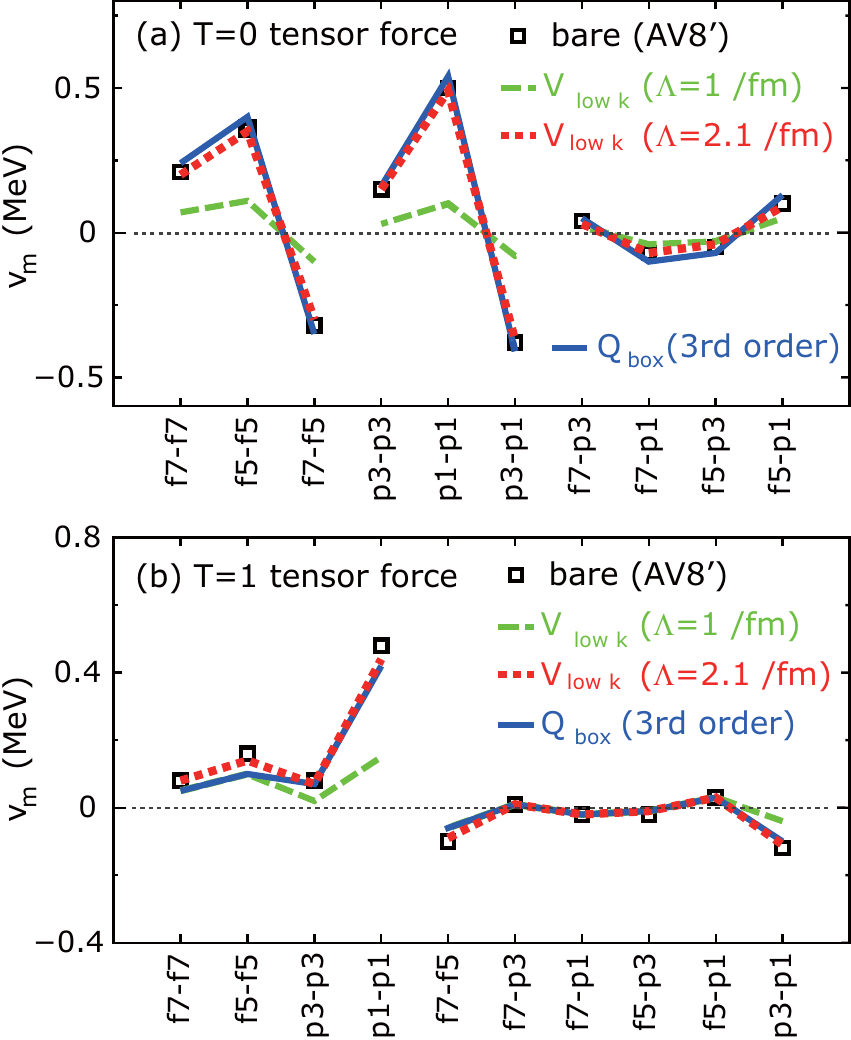}
 \caption{From \cite{otsuka}: Tensor forces in AV8' interaction, in low-momentum interactions in the pf shell
obtained from AV8', and in the 3rd-order Q$_{box}$ interaction for (a) T=0 and
(b) T=1. Results in S$n$EFT treated in $V_{lowk}$ are similar. The Q$_{box}$ contains higher-order $1/\bar{N}$ corrections in ring-diagram approximation. 
}
  \label{pfshell}
 \end{center}
\end{figure}

This leads to the next observation.
\vskip 0.3cm
$\bullet$ {\bf  The nuclear tensor force derived from the $V_{lowk}$-RG is a Landau Fermi-liquid fixed point interaction.}

The smoothed cusp structure in $E_{sym}$ near $n_{1/2}$ gives rise to the changeover from softness to hardness in the EoS mentioned above. Its precise form depends on where the density $n_{1/2}$ is located. Given that its precise location cannot be given by the theory\footnote{More on this point below in connection with what happens above the density $n_{1/2}$.}, there is an unavoidable uncertainty in certain astrophysical properties (such as the tidal deformability $\Lambda_{1.4}$ observed in the gravity-wave measurements which is highly sensitive to density) that cannot be precisely pinned down by theory~\cite{PCM}. A highly important point to underline at this point is that given the future precision measurements to be done for the density ladder near nuclear matter density~\cite{SP-MDPI},  the $V_{lowk}^{\rm tensor}$,  a Fermi-liquid fixed point quantity,  is poised  to play an extremely important role in the EoS. As shown in \cite{nuclear-cusp}, it reproduces the inflection smoothed from the cusp for $E_{sym}$ embodied in the topology change hidden in nuclear correlations.  
\section{Pseudo-conformal (PC) phase}
Up to the onset of the half-skyrmion phase, the parameters that enter in $\cal{L}_{\psi {\cal V} \chi}$ remain unchanged from the BR scaling as originally proposed. However they get modified significantly  by the topology change at $n\gsim n_{1/2}$. The principal change is the HLS gauge coupling $g_{hls}$ that no longer is $U(2)$ symmetric, with $g_\rho$ and $g_\omega$ behaving differently as density increases above $n_{1/2}$. The most crucial effect is the $SU(2)$ gauge coupling $g_\rho$ which flows toward the VM fixed point (VMFP)~\cite{HY-HLS} $g_\rho \to 0$ at high density $\gsim 20 n_0$ whereas $g_\omega$ deviates from the flow to the VMFP in close association to the dilaton-nucleon coupling $g_{\chi N}$ required for the stability of the matter~\cite{PKLMR}.  Certain topological properties of the half-skyrmion structure in the phase $n\gsim n_{1/2}$, such as the emergence of baryon parity-doubling symmetry, the approach to the dilaton fixed point with $f_\chi\to f_\pi$, $g_A\to 1$ etc. enter in the parameters of the Lagrangian $\cal{L}_{\psi {\cal V} \chi}$. Not rigorously formulated though they were, they constrain in a way more or less to globally control the properties of massive compact stars in agreement with the available astrophysical observations. 

We will skip the details which are reviewed in \cite{PCM} and {\it focus on one phenomenon on which Tom Kuo's role figures crucially}.  It is a unique and unambiguous prediction  in G$n$EFT that differs from all other predictions found in the literature. It deals with the sound speed $v_s$ in the core of massive stars  with the baryonic matter ranging in densities from $\sim 3n_0$ to $\sim 7n_0$ arising  from hidden local symmetry (HLS) with the gauge coupling $g_\rho$ flowing  to the vector manifestation (VM) fixed point at high density. 
The density at which the VM fixed-point is located cannot be pinned down by theory:  lattice QCD measurement feasible in temperature cannot access high densities. We consider two densities, one ``low" density,  $n_{core}\sim 6 n_0$, and the other ``high" density,  $n_{asymp}\gsim 25 n_0$,  that we consider high enough to be ``asymptotic."  The density $n_{core}$ is considered commonly as the density at which quarks could get ``deconfined."  We show that while most of the global star properties are not sensitively dependent on where the VM fixed point is located,  there is one star property that our approach distinguishes at what VM density the pseudo-conformaliy can emerge in dense medium.
\subsection{Half-skyrmion phase}
It was mentioned that viewed from the chiral symmetry point of view, the half-skyrmion phase arising at higher density differs from the skyrmion phase in that the quark condensate averaged on the crystal lattice goes to zero but the pion decay constant does not, so chiral symmetry remains in the NG mode. This ``pseudo-gap"-like phenomenon impacts how the symmetry energy is tied to the tensor-force structure as density goes from below to above $n_{1/2}$. But how the sound speed  behaves in the half-skyrmion phase  could not be directly linked to the pseudo-gap property. It instead reflects how hidden scale symmetry emerges due to strong correlations in the higher density 1/2-skyrmion regime. 
To explore this aspect, we examined how the half-skyrmion phase in the skyrmion-crystal simulation of dense baryonic matter looked like~\cite{PKLMR}.  The result was surprising: What happened closely resembled a Landau-Fermi liquid fixed point where the $\beta$ function for the quasiparticle interactions is entirely suppressed. 

\vskip 0.4cm

$\bullet$ {\bf Emergent scale symmetry for $\mathbf{n > n_{1/2}}$}:
%
We think it deserves a bit of detail given in Ref.~ \refcite{PKLMR} for understanding the pseudo-conformal structure at density $\gsim n_{1/2}$. Write the chiral field $U$ as $U(\vec{x}) = \phi_0(x,\, y,\, z ) + i \phi^j_\pi(x,\, y,\, z )\tau^j$  with the Pauli matrix $\tau^j$ and $j=1,2,3$. Including the $\rho$ and $\omega$, we write the fields placed in the lattice size $L$ as $\phi_{\eta,\, L}(\vec{x}\,)$ with $\eta =0,\, \pi,\, \rho,\, \omega$ and normalize them with respect to their maximum values denoted $\phi_{\eta,L,{\rm max}}$ for given $L$.  It comes out with the mesonic chiral-scale symmetric Lagrangian $\cal{L}_{ {\cal V} \chi}$ -- namely, $\cal{L}_{\psi {\cal V} \chi}$ without $\psi$ but with the homogeneous  Wess-Zumino term~\cite{HY-HLS} implemented for $N_f=2$ -- in the half-skyrmion phase with $L\lsim L_{1/2}$ where  $L_{1/2}\simeq2.9$ fm, that the field configurations are invariant under scaling in density as the lattice is scaled from $L_1$ to $L_2$.
\be
\frac{\phi_{\eta,\,L_1}(L_1\vec{t}\,)}{\phi_{\eta,\,L_1,\,{\rm max}}}= \frac{\phi_{\eta,\,L_2}(L_2\vec{t}\,)}{\phi_{\eta,\,L_2,\,{\rm max}}}. \nonumber
\ee
\begin{figure}[h]
\begin{center}
\includegraphics[width=5.2cm]{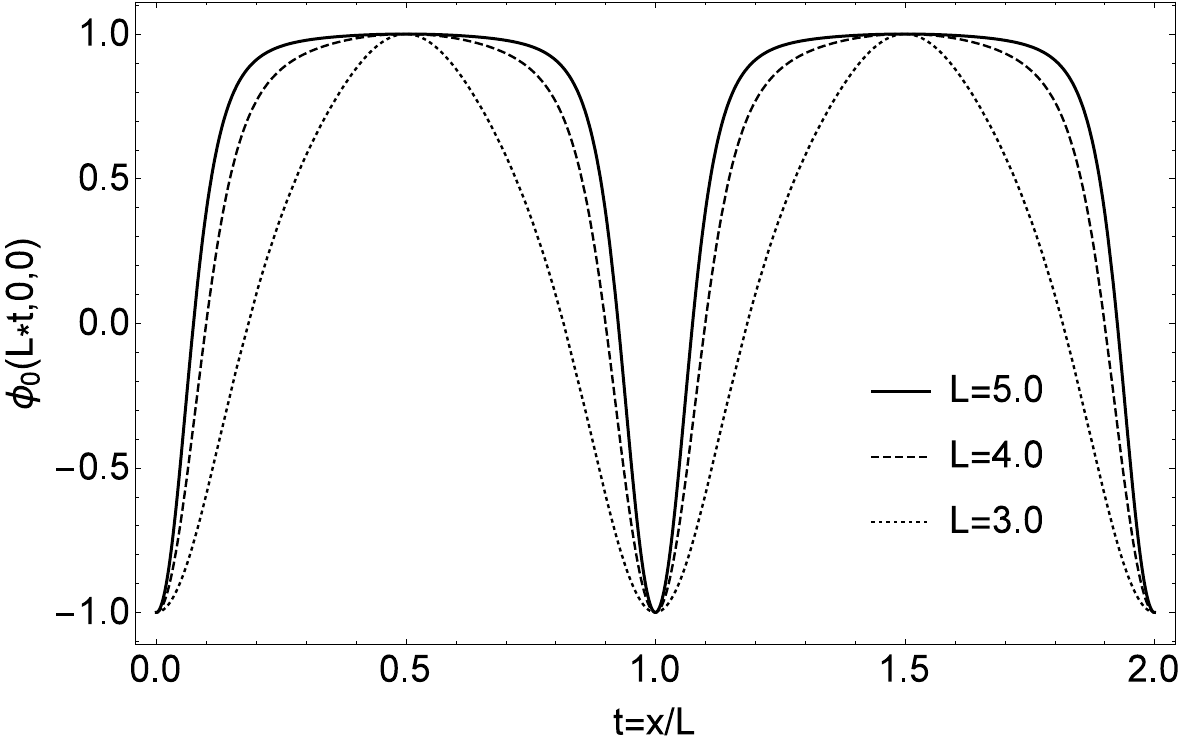}\includegraphics[width=5.2cm]{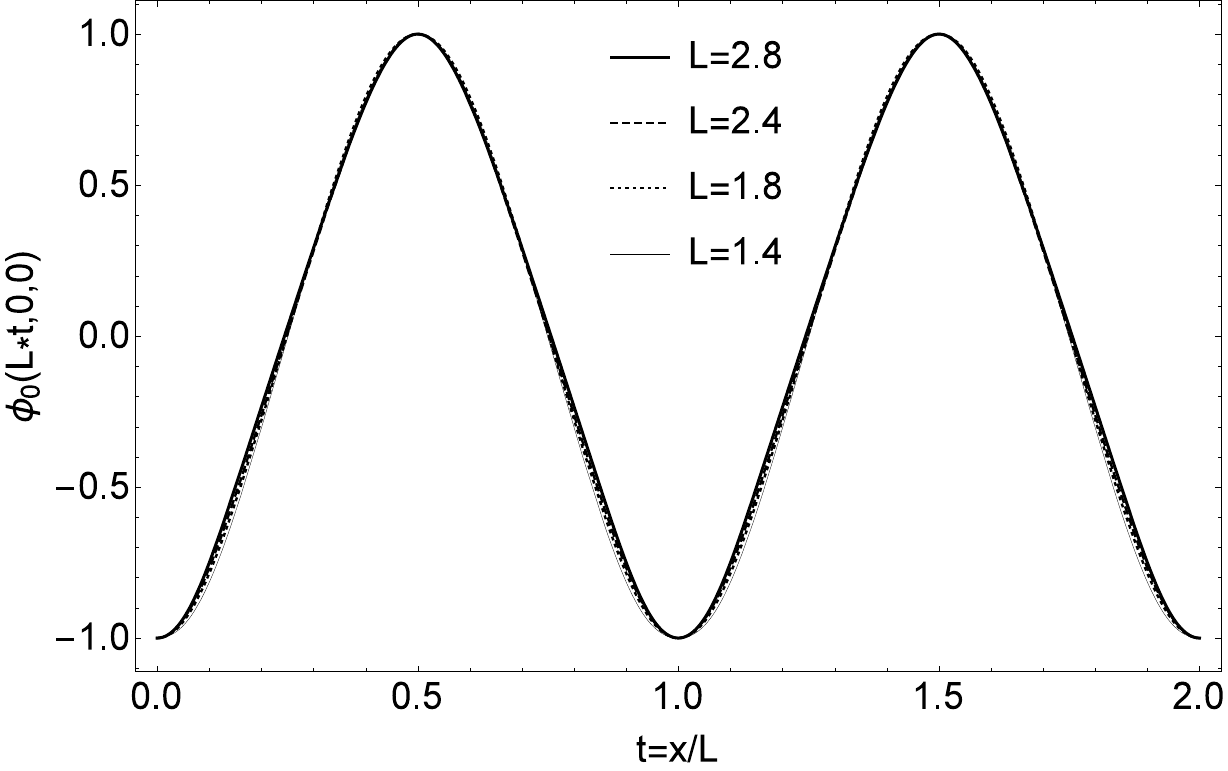}
\includegraphics[width=5.2cm]{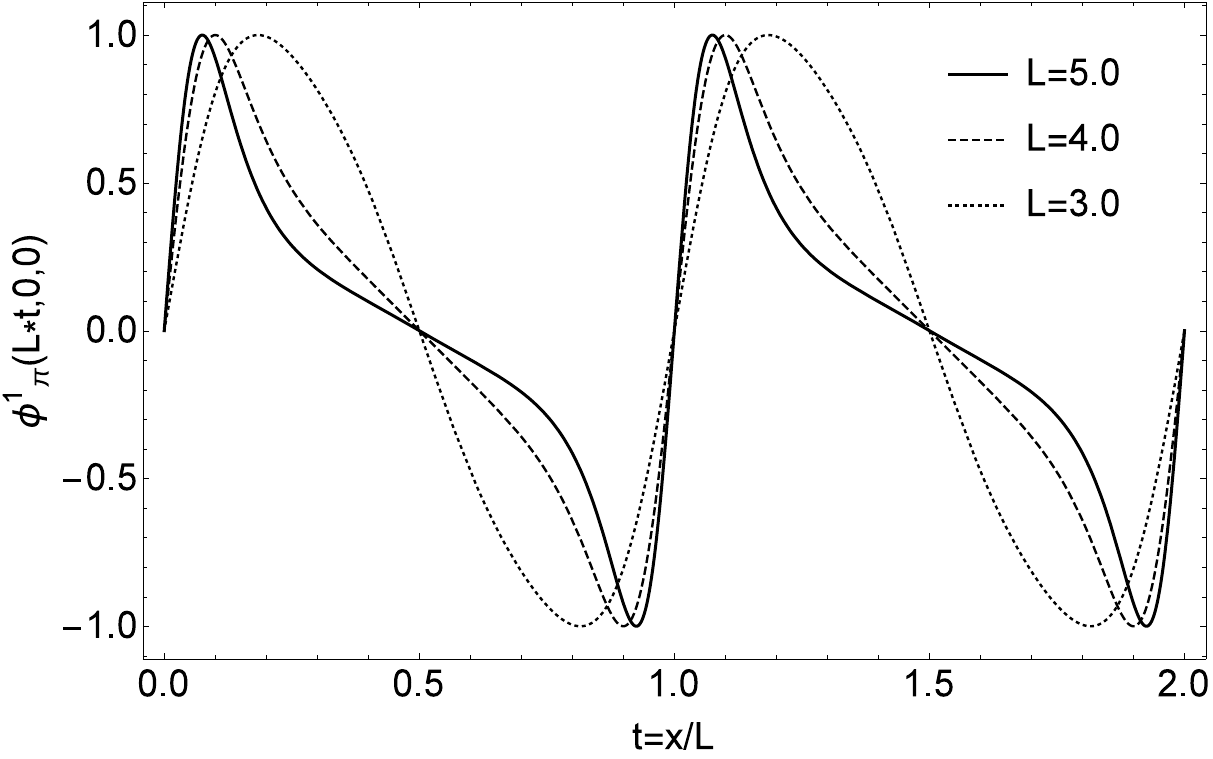}\includegraphics[width=5.2cm]{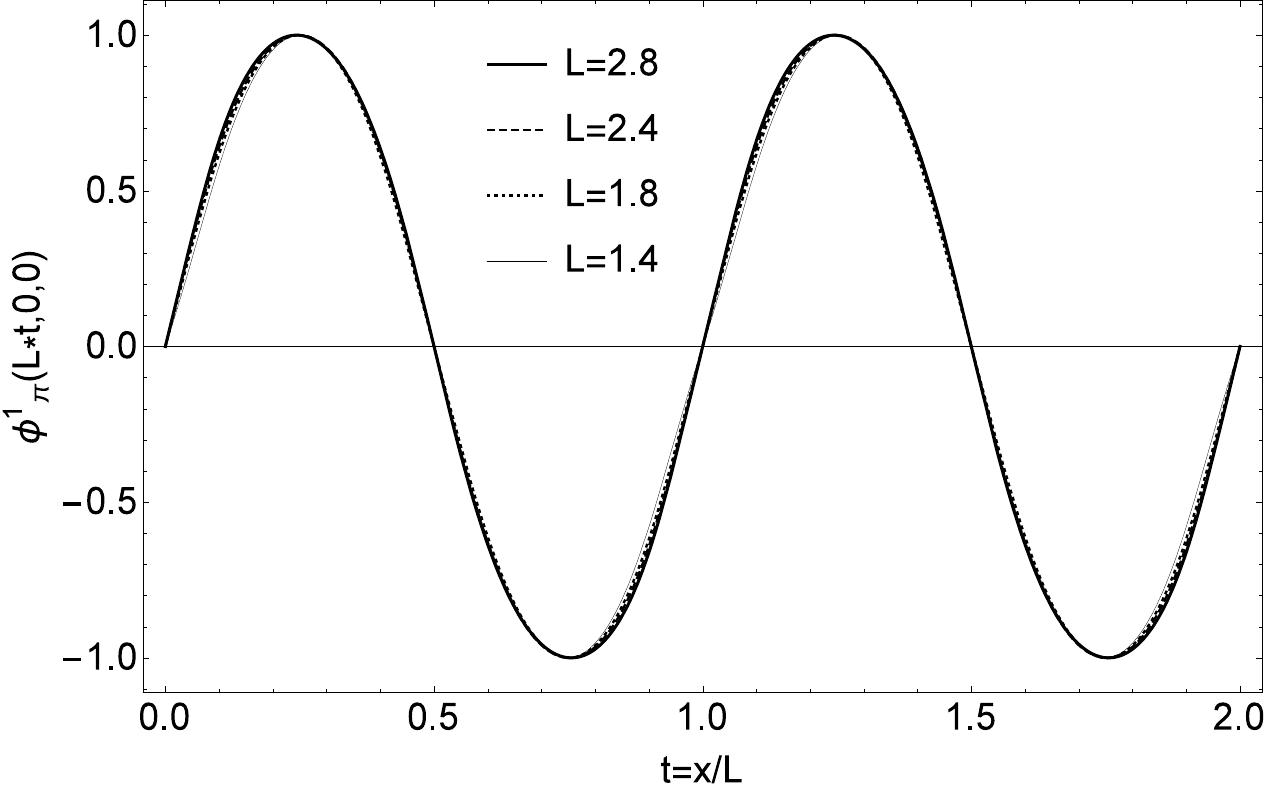}
\caption{  The field configurations $\phi_0$ and $\phi^1_\pi$  as a function of $t = x/L$ along the y = z = 0 line. The maximum values for $\eta=0,\pi$ are $\phi_{0,\,L,\, {\rm max}} = \phi_{\pi,\,L,\, {\rm max}} = 1$. The half-skyrmion phase sets in when $L=L_{1/2} \lsim 2.9\,{\rm fm}$.
 }\label{scale_inv}
\end{center}
\end{figure}
The crystal lattice simulation is shown in  Fig.~\ref{scale_inv} for  $\phi_{0,\pi} (t, 0,0)$ vs. $t$ with $t\equiv x/L$~\cite{PKLMR}.  Other fields behave similarly.  One can see that in the skyrmion phase with lower density with $L >  L_{1/2}$, the field is clearly -- as expected -- dependent on density. But it is markedly independent of density in the half-skyrmion phase with $L\lsim L_{1/2}$.

It is worth pointing out an intriguing observation here although one cannot at present give a convincing argument. What this result shows is that deep in the half-skyrmion phase, scale invariance could set in even though the trace of the energy-momentum tensor  is not equal to zero.  This may be taken a signal for the emergence of pseudo-conformality.  It is tempting to interpret this a  skyrmion matter arising from the ``conformal pions"  in deep infrared region for the quark mass anomalous dimension $\gamma_*=1$~\cite{zwicky,beta*}. There the composite pionic structure in the kinetic energy term for the HLS $\rho$ field as an interpolating field could stabilize the skyrmion in dense matter like the Skyrme quartic term does as conjectured by Yamawaki~\cite{HLS-grassmannian}.  Of course one may not rely on this picture near $n_{1/2}\approx n_{hqc}$ -- perhaps it is too low -- but it is plausible in the core of massive neutron stars.  We will come back to this feature below in connection with the Fermi-liquid fixed-point $g_A^L=1$ in nuclear Gamow-Teller superallowed transitions.
\subsection{``Pseudo-conformal" sound velocity $v_{pcs}$}
In the low density phase with $n\lsim n_{1/2}$, the parameters of the Lagrangian $\cal{L}_{\psi {\cal V} \chi}$ are given by the BR scaling as noted above. There the star (static) properties such as the symmetry energy $E_{sym}(n)$ etc. will be those of an improved CDF approach refined by S$n$EFT for any value of $n_{1/2}$ . Derivatives with respect to density of the symmetry energy will however depend on the location of $n_{1/2}$. 

Going above the topology-change density $n_{1/2}$, due to various interplays of quasi-nucleon mass, scaling coupling constants, vector repulsion etc. figuring in the EoS,   the topology-change density $n_{1/2} \gsim 4 n_0$ giving $M_{max}\gsim 2.4 M_\odot$ seems to be ruled out by causality. Otherwise the ranges $2\leq n_{1/2}/n_0  < 4$ yield similar global star properties. Our discussion will be only for the $n_{1/2}\sim 2.0 n_0$ predictions.\footnote{In the reviews~\cite{PCM}, the favored density for $n_{1/2}$ was $n_{1/2}\sim 3.5 n_0$.}

The mean-field treatment of the Lagrangian $\cal{L}_{\psi {\cal V} \chi}$ (equivalently the Fermi-liquid fixed-point approximation in G$n$EFT) gives the VeV of the TEMT (in the chiral limit)  $\theta_\mu^\mu$~\cite{PKLMR}
\begin{eqnarray}
\langle \theta^\mu_\mu \rangle
 = 4V(\langle \chi \rangle) - \langle \chi \rangle \left. \frac{\partial V( \chi)}{\partial \chi} \right|_{\chi = \langle \chi \rangle}
\ee
where $V(\chi)$ is the dilaton potential.  It can be seen that the Fermi surface does not spoil  the scale symmetry. Up to the density $n_{1/2}$, this is more or less what's given by S$n$EFT. However most surprisingly a drastic difference arises to the sound velocity $v_s$ depending on whether the VM fixed point lies low or high even though  global star properties seem to stay more or less the same.  This is seen in Fig.~\ref{sound}. With the changes in the parameters going across $n_{1/2}$\footnote{The parameterization made in Ref.~\refcite{PKLMR} looks somewhat complicated but can actually be made much simpler without changing the essential feature as the crystal simulation indicates. What's shown in the reviews~\cite{PCM} contains this improvement.} duly taken into account, the sound velocity  computed in the $V_{lowk}$-RG comes qualitatively differently.
\begin{figure}[h]
\begin{center}
\includegraphics[width=9.7cm,angle=0]{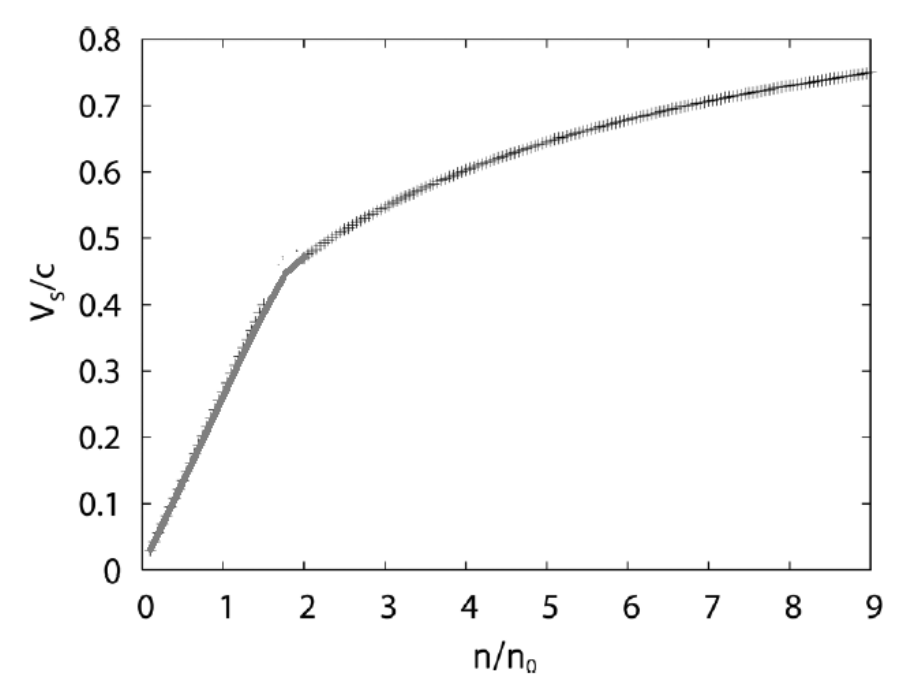}
\includegraphics[width=9cm]{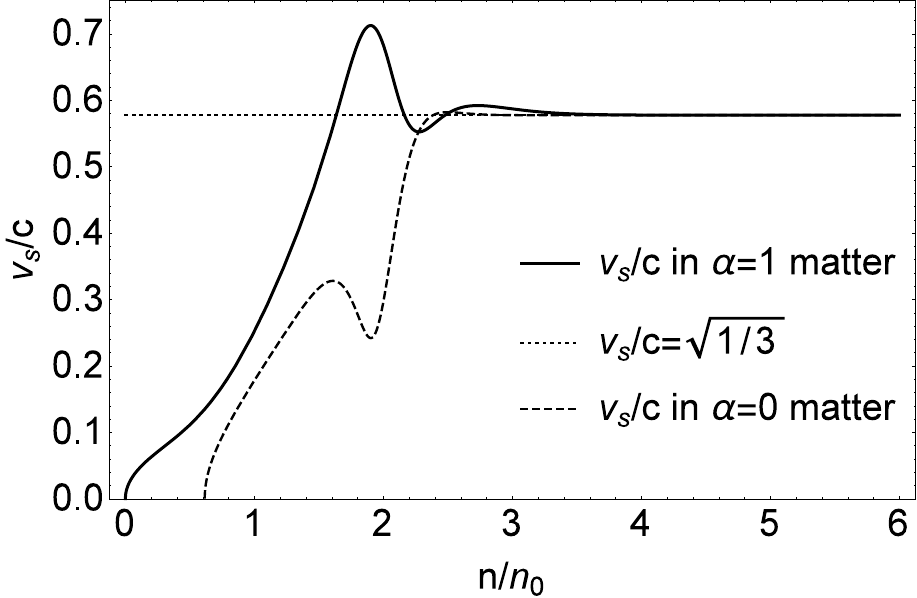}\caption{ The sound velocity $v_{pcs}$ predicted in $V_{lowk}$-RG for $n_{VM}=6n_0$ (upper panel, neutron matter)~\cite{PKLR} and for $n_{VM}=25 n_0$ (lower panel))~\cite{PKLMR}. For illustration the topology change density was chosen for $n_{1/2}=2n_0$.}
\label{sound}
\end{center}
\end{figure}
The upper panel is for a low VM density, say,  $n_{VM}=6n_0$, a density often considered appropriate for the star core in models with deconfined quarks. It reproduces the structure for the stars considered populated by standard nucleons as exemplified by the (well-known) Akmal-Pandharipande-Ravenhall  nuclear EoS~\cite{APR}.  Although it violates causality at high density it exemplifies the class of hadronic-star structure more or less consistent with the current developments of astrophysical observations~\cite{weise}.

The lower panel shows an unexpected feature. For what might be considered as ``asymptotically high density" $n_{VM}=25n_0$, we observe a structure distinctly different from the usual hadronic EoS. It shows what we refer to as ``pseudo-conformal" sound velocity $v_{pcs}^2/c^2\approx 1/3$ converged to slightly above the topology change density. It is  preceded by a bump exceeding the conformal velocity 1/3.  It is not clear what produces this bump but it seems to be correlated with the nature of the overlap between, in skyrmion picture, the crossover from {\it baryonic} structure to {\it quarkish} structure, capturing the putative ``hadron-quark duality," resembling what takes place in  quarkyonic descriptions~\cite{quarkyonic}. The sound speed is not conformal in this PCM because the TEMT is not equal to zero because the scale symmetry is not fully -- but only  approximately -- restored there. The non-zero TEMT is density-independent, however, so its derivative with respect to density is equal to zero revealing $(v_s^2 -1/3)\to 0$. It is most likely not strictly zero because there can be correction terms that cannot be ignored completely.   As will be pointed out below, the predicted properties are much like those of the deconfined quarks. This feature can be understood in terms of the hidden symmetries involved and the Cheshire Cat mechanism, viz, Cheshire Catism~\cite{CC1,CC2}.
We further note that this strikingly different sound speed could very well be related to  the appearance of half-skyrmion configurations  at some high density above $n_{1/2}$ seen in Fig.~\ref{scale_inv}. One may understand why $n_{VM} << 25 n_0$, e.g., the upper panel of Fig.~\ref{sound}, does not favor the onset of emergent scale symmetry. 

As stated below, it has been an intriguing question to the aficionados of the PCM as to whether there is no ``smoking gun" constraint to rule out one of the two qualitatively different scenarios of Fig.~\ref{sound} in upcoming astrophysical observables.  Detailed analyses of the up-to-date available observables~\cite{weise} indicate that both of the two scenarios as they stand cannot yet be ruled out. 
%
%

\subsection{Parameterizing pseudo-conformality}
Given the pseudo-conformal sound velocity calculated in the $V_{lowk}$-RG with the parameter changes at $n_{1/2}$ taken into account in the Lagrangian $\cal{L}_{\psi {\cal V} \chi}$, it is feasible to reproduce the result by simply replacing the EoS for $n > n_{1/2}$ by parameterizing the single-particle energy of the matter $E_0/A$ that satisfies $\frac{d\, P}{dn} = \frac{1}{3} \frac{d\,\epsilon}{dn}$ -- which assumes the density independent TEMT in $n > n_{1/2}$~\cite{PKLMR}
\begin{equation}
E_0/A = -m_N + B \left( \frac{n}{n_0} \right)^{1/3} + D\left( \frac{n}{n_0} \right)^{-1}. \label{EoSRII}
\end{equation}
One can do this replacement for the range of $2.0 \lsim n_{1/2}/n_0 <4.0$,  accurately reproducing $v_{pcs}$ computed by $V_{lowk}$\footnote{The parameters $B$ and $D$ are to be fixed at $n=n_{1/2}$ by continuity in the chemical potential and pressure.  For $n_{1/2}=2.0n_0$, the parameters $B_{\alpha=(0,1)}= (570 \textnormal{ MeV},\, 686 \textnormal{ MeV})$ and  $D_{\alpha=(0,1)}= (440 \textnormal{ MeV},\, 253\textnormal{ MeV})$ the reproduction is almost exact.} that simplifies the calculation of $v_{pcs}$ for any given $n_{1/2}$ in the range $2.0 \lsim n_{1/2}/n_0 <4.0$.  An interesting observation is that if one takes  $n_{1/2} \geq 4.0 n_0$, then causality is violated as mentioned. Naively this would imply that the maximum star mass accommodated in the PCM is $M_{max}\sim 2.4 M_\odot$.  Whether or not such high mass stars can still be accommodated by fiddling the parameters of the Lagrangian  $\cal{L}_{\psi {\cal V} \chi}$ without bringing tension to other observables has not yet been investigated. 

An important remark to make at this point is that {\t it is essentially the pseudo-conformality symmetry that seems to control the sound speed $v_{psc}$  from $n_{1/2}$ on,  embodying what appears to be rather complicated interplays that go into the parameters of the Lagrangian.}  We take this as a support for the simplification  made in the parameters based on topological properties. But it remains unclear as to what the basic physics mechanism is in leading -- or not -- to the pseudo-conformal sound speed, with qualitatively different dependence on  the vector manifestation density ($n_{VM})$ while other star properties are little affected by the location of $n_{VM}$. 
\section{Fermi-Liquid Fixed Point $\mathbf{g_A^{\rm L}}=1$}
Not directly connected, but implicated with the underlying hidden symmetries, to the sound velocity is the $g_A$ problem in nuclear beta decay. It has been argued since some time that what seems to be observed in allowed -- more strikingly in superallowed --  Gamow-Teller transitions in nuclei requires an effective axial coupling constant $g_A^\ast\simeq 1$~\cite{gA-review} be identified with a Fermi-liquid fixed point in the $\sigma\tau$ channel~\cite{FR,gAMR}. If one ignores the possible effect of the trace anomaly discussed in \cite{anomaly-induced}, this phenomenon could be systematically analyzed in the $V_{lowk}$-RG approach.  

What is involved is the nuclear matrix element of the  axial current
\be
J^{a\mu}_{ 5}=g_A  \bar{\psi}\gamma^\mu\gamma_5 \frac{\tau^a}{2}\psi. \label{AC}
\ee
In the GD scheme of scale symmetry \`a la Crewther~\cite{GD}, there is an anomaly-induced factor dependent on the anomalous dimension $\beta^\prime$\footnote{It is interesting that this effect is induced by nuclear process, not present in the vacuum, although the effect itself is related to a fundamental anomaly in gauge theory.}. It will be ignored as mentioned.\footnote{In  Zwicky's QCD-CD scheme~\cite{beta*}, $\beta^\prime=0$, so this factor is absent. Higher chiral-scale terms could enter in its place and need to be looked at~\cite{LQinprogress}.}  

The current (\ref{AC}) as written is scale-invariant. Coupled to quasiparticles on the Fermi surface, it will excite from the $J=T=0$ ground state a $J=T=1$ quasi-particle-quasi-hole-state. The quasi-p(article)-quasi-h(ole) states interact via the Landau(-Migdal) Fermi-liquid fixed point interaction $G_0^\prime$.\footnote{The pion field enters in the theory so it should be referred to as Migdal theory~\cite{migdal}.} The quasi-p-quasi-h bubbles are suppressed by the $1/\bar{N}$ factors\footnote{The Landau-Migdal $G_0$ interaction is supposed to be marginal like other Landau parameters}, so the full matrix element of the current  gives rise to the Fermi-liquid fixed point quantity $g_A^{\rm L}$ multiplying the non-interacting quasi-p-quasi-h matrix element $\la ph|\sigma\tau|0\ra$
\be
J^{\rm GT}=g^{\rm L}_A (\sigma\tau)_Q
\ee 
with the subscript $Q$ indicating that $\sigma\tau$ operates on non-interacting single-quasi-p-quasi-h states.

How to determine theoretically the FLFP quantity $g_A^{\rm L}$ was suggested by the Shankar-Polchinski approach applied to the hidden symmetry Lagrangian in \cite{FR}. There nuclear EW response functions at low-momentum transfer processes were very accurately reproduced by the mean-field approximation -- corresponding to the Fermi-liquid fixed-point approximation as noted -- of Lagrangian $\cal{L}_{\psi {\cal V} \chi}$ with the BR scaling encoded. For the EM, it was beautifully illustrated with the proton anomalous gyromagnetic ratio in Pb nuclei $\delta g_l^p$ (which agreed very closely with the experimental value).\footnote{As far as we are aware, there are no other nuclear physics calculations, Landau-Migdal or shell-model or otherwise, that  match this prediction in precision.} The same approximation led to
\be
g_A^{\rm L}=g_A q^{\rm L}_{snc}
\ee
with
\be
q^{\rm L}_{snc}=(1- \frac 13 \Phi^\ast \bar{F}_1^\pi)^{-2}.\label{snc}
\ee
where
\be
\Phi^\ast=f_\chi^\ast/f_\chi\simeq f_\pi^\ast/f_\pi
\ee
and $\bar{F}_1^\pi$ is the Landau-Migdal pionic interaction parameter which could be taken as a Fermi-liquid fixed-point parameter for given Fermi momentum on the Fermi surface. The product $ \Phi^\ast \bar{F}_1^\pi$ controlled by chiral symmetry turns out to be little sensitive to density,  varying only a few \% between $1/2 \lsim n/n_0\lsim 1$, hence gives 
$q^{\rm L}_{snc}\simeq 0.78$ leading to
\be
g_A^{\rm L}\simeq 1.0\label{gA=1}
\ee
for light as well as heavy nuclei. 

The result (\ref{gA=1}) has been known since many years. But there was one assumption which was not fully justified then. It was the Goldberger-Treiman relation connecting $g_A$ to the $g_{\pi NN}$ coupling valid for large $N_c$ in the skyrmion description where the $g_{\pi NN}$ coupling was identified with the Skyrme quartic term that stabilizes the soliton. In the new formulation of the HLS $\rho$ meson as a dynamical gauge boson in the grassmanian manifold~\cite{HLS-grassmannian} the Skyrme term figures as the {\it kinetic energy term} for the composite field for the $\rho$ meson. This identification was mentioned above on scale-invariant half-skyrmions in the deep infrared regime supporting non-interacting pion fields.

There is a support for (\ref{gA=1}) in the superallowed Gamow-Teller transition in the doubly-closed shell nucleus $^{100}$Sn in a measurement at GSI~\cite{GSI}. But a more recent, what's heralded more precise, measurement at RIKEN in the same nucleus~\cite{Riken} seems to {\it strongly} disagree with the GSI result. If the RIKEN result is confirmed, it will raise an extremely serious issue totally unrecognized in the field in the ongoing experiments on $0\nu\beta\beta$ transitions~\cite{anomaly-induced}. While it does not seem to concern the $g_A^{\rm L}$ problem, it will be a serious issue for the physics BSM.

It is worth noting that there is an indication that the possible correction to (\ref{gA=1}) could be quite negligible. The prediction (\ref{gA=1}) is obtained at the Fermi-liquid fixed point approximation corresponding to $1/\bar{N}\approx 0$. A recent calculation using the coadjoint-orbits approach~\cite{dtson} of the leading  $1/\bar{N}\approx 0$  correction indicates that the corrections can indeed be negligible, $\sim O(10^{-4})$~\cite{shao}.

Finally a remark as to why this issue of $g_A^{\rm L}$ has anything to do with the PC sound speed.  

In  finite nuclei, the effective $g_A\simeq 1$ seen in simple shell-model description permeates from light to heavy nuclei as $g_A^{\rm L}\simeq 1$. Now if one takes what's called ``dilaton-limit fixed-point (DLFP)"~\cite{DLFP} on the mean-field of the Lagrangian $\cal{L}_{\psi {\cal V} \chi}$,
\be
{\rm Tr} (\Sigma\Sigma^\dagger)\to 0
\ee 
where $\Sigma=U \chi(f_\pi/f_\chi)$, 
there arise singularities. To avoid these singularities for consistency of the theory, here G$n$EFT, we are required to impose the ``dilaton-limit fixed-point" constraint
\be
g_A=g_V \to 1,\ f_\pi\to f_\chi
\ee
leading to a sigma-model-type Lagrangian with the $\omega$ gauge coupled to baryons and pions with, however,  the $\rho$ meson decoupled.

There is no way known to determine what the DLFP density is, in particular, in relation to the VM fixed-point, the emergence of parity-doubling in the nucleon structure, etc. What is however notable is that $g_A=1$ seems to permeate from low density to high density in a way similar to the pseudo-conformality as in the sound velocity. This phenomenon, if confirmed, would set a new paradigm in nuclear physics.

\section{Further remarks}
We have not discussed the possibility of going top-down the EoS ladder in pQCD. The question is how perturbative QCD builds the ``architecture" to enable one to go down  from asymptotic density to S$\chi$EFT and ask where the ladder step can be made to meet the going-up step at G$n$EFT.  There is at present just as deep a ``jungle" in going top-down as there is in  bottom-up, so there has been no clear inkling of what's actually happening~\cite{pQCD-jungle}. 

Let us briefly compare what we predict in the PCM with what's considered as ``evidence" in quark-matter core~\cite{annala}. The PCM predictions~\cite{CC2} are that for a star of $M\sim 2.3 M_\odot$, one should expect $v^2_s/c^2 \approx 1/3$, $\Delta\equiv 1/3-P/\epsilon\approx 0.08$ and the polytropic index $\gamma ={\rm d\ ln} p/{\rm d\ ln}\epsilon\to 1$ inside the core of massive stars. These are characteristic features of ``deconfined quarks" offered in articles citing evidence for quark matter~\cite{annala}. But the matter in the PCM is of hadronic quasiparticles, not of ``deconfined quarks."  One can however speculate certain scenarios on how quasi-baryons can masquerade fractionally charged objects~\cite{speculations}.  One such scenario is a sort of Cheshire Cat mechanism~\cite{CC1} involving  domain walls  as in  condensed matter systems which support deconfined quasiparticles on the stack of sheets. It could possibly pass a class of  ``Duck Test."\footnote{ Duck Test: ``If it looks like a duck, walks like a duck and quacks like a duck, then it just may be a duck."} It would of course be a challenge in nuclear physics to come up with convincing explanations: It must involve an extremely intricate interplay in G$n$EFT between the $(\rho, \omega)$-nuclear coupling or decoupling and the dilaton condensate together with the emergence of parity-doubling in the nucleon spectrum~\cite{PKLMR,WCU}. 
\subsection*{Acknowledgments}
One of the authors (MR) is grateful for useful discussions via E-mails with Hyun Kyu Lee and Wolfram Weise.  The other acknowledges helpful comments from  Yong-Liang Ma.
 
%
%
%

%
\end{document}